\documentclass[11pt]{JHEP3}
\usepackage{graphicx,amssymb}
\usepackage{epsfig}

\newcommand{\newc}{\newcommand}
\newc{\eeq}{\end{equation}}
\newc{\beq}{\begin{equation}}
\newc{\ec}{\end{center}}
\newc{\bc}{\begin{center}}
\newc{\eeqa}{\end{eqnarray}}
\newc{\beqa}{\begin{eqnarray}}
\newc{\nonr}{\nonumber}
\newc{\bd}{\begin{description}}
\newc{\ed}{\end{description}}
\newc{\benu}{\begin{enumerate}}
\newc{\eenu}{\end{enumerate}}
\newc{\bi}{\begin{itemize}}
\newc{\ei}{\end{itemize}}
\newc{\ra}{\rightarrow}
\newc{\bis}{\begin{itemstep}}
\newc{\eis}{\end{itemstep}}

\newc{\LH}{\hat{L}}
\newc{\RH}{\hat{R}}
\newc{\SL}{\not\!}
\newc{\mtr}{\mathrm {tr}}

\title{An Effective Operators Analysis of Leptonic CP Violation :
Bridging High and Low Energy Processes }
\author{We-Fu Chang\\Institute of Physics, Academia Sinica, Nankang, Taipei, Taiwan 115\\
E-mail: \email{wfchang@phys.sinica.edu.tw}}
\author{John N. Ng\\ Theory Group,
TRIUMF, 4004 Wesbrook Mall,
Vancouver, B.C.,
Canada V6T 2A3\\ E-mail: \email{misery@triumf.ca}}

\abstract{
We study  the leptonic CP violation by employing the
complete set of dimension-six pure leptonic effective operators.
Connection among the observable at different energy scales can be
made by the running of the renormalization group equations.
Explicitly, we study the  charged lepton electric dipole moment, muon Michel decay,
and the triple spin-momentum correlations at the Linear Collider.
We found the electron electric dipole moment, which starts at
2-loop level, severely constrains the possibilities to detect the
CP violating signatures in muon decay and at the linear colliders.
}

\begin{document}
\section{Introduction}
The search for time reversal odd (TO) correlations in charged current decays
of nuclei and mesons such as the kaon and the muon has a long and venerable history.
Parallel to these are the ongoing experiments in search of TO signatures
in flavor conserving interactions highlighted by the many electric dipole
moment (EDM) searches. Within the standard model (SM) occurrence of such signals
are highly forbidden. This is because  the SM contains only one CP violating source
 which is the Cabibbo-Kobayashi-Maskawa phase in the quark mixing matrix.
The lowest order CP violating or TO effects are associated with flavor changing
neutral currents in the quark sector. Charged current and flavor conserving TO are
highly suppressed to three loops or higher.  However, many models of physics
beyond the SM have more sources of CP violation and such experiments take on
the role of probing new physics. The seemingly countless number of new physics
models all have to meet the constraints arising from the search of the permanent
 electric dipole moment (EDM) of the electron and now serves
as the most stringent and theoretically very clean test of these models.
A review of the literature can be found in  Ref.(\cite{edmth}).
For a more up to date discussion that connects with  neutrino mass generation
see Ref.(\cite{CN}) and split supersymmetry see Ref.(\cite{AD,CCK}). In this
paper we explore the relationship  between T-violating neutral current (TVNC)
and  the EDM of charged leptons; in particular that of the electron
$d_e$ and the muon $d_\mu$ which will be probed to unprecedented accuracy
in the next generation of dedicated experiments.

Our analysis employs an effective field theory approach as oppose to focusing
on a  particular model. We first assume that new physics occurs at a scale $\Lambda$
above the weak scale. We can take this to be the mass scale of the new degrees
of freedom that have been integrated out. Their manifestation at energies below
$\Lambda$ will be a set of  operators made up of SM fields and they have dimensions
higher than four. A list of such operators are given in \cite{oplist}. Dimension
six operators that violates baryon and lepton numbers were given earlier in \cite{opbv}.
We shall ignore the latter and assume the proton to be stable.
 From dimensional considerations the
leading contributors will be  of dimension six\footnote{The lone dimension
five operator is related to neutrino masses and does not enter our discussions.}
as they are  suppressed by $1/\Lambda^2$. Higher dimensional operators are
suppressed by higher powers of $1/\Lambda$. Specifically, we focus on dimension six
four fermion operators that contains leptons only.
Between the electroweak scale characterized by  $v\simeq 250$ GeV  and
$\Lambda$ which we take to be above 1 TeV the gauge symmetry and particle content
are that of the SM. The fermions are all chiral at this stage. As will be
seen in greater detail below new charged currents will appear.
Their Lorentz structures are restricted by $SU(2)\times U(1)$ gauge symmetry
and chirality.
At the weak scale spontaneous symmetry breaking takes place and the fermions
and weak bosons become massive. Since we do not discuss neutrino masses and their
effects the addition of right-handed singlets is not mandatory here.
However, we do take them to be massive. Below $\Lambda$ the effective Lagrangian
consists of the SM plus a sum of dimension six operators multiplied
by their Wilson coefficients.
Renormalization will mix these operators and the renormalization group
equations (RGE) which will be derived are used to run the Wilson coefficients
to the electroweak scale. Below the weak scale the RG running of the SM will take place.
The EDM's are generated at two loops from the SM terms and the new operators.
This is how the TVNC and the EDM connection is established.
A necessary condition is that not all the Wilson coefficients can be made real.
We shall show that if there is no fine tuning of parameters  or setting phases
to zero by hand this is indeed the case in general.
Equipped with this and standard effective field theory techniques we
obtained constraints on these possible TVNC interactions.

\begin{figure}[h]
      \begin{center}
    \epsfig{file=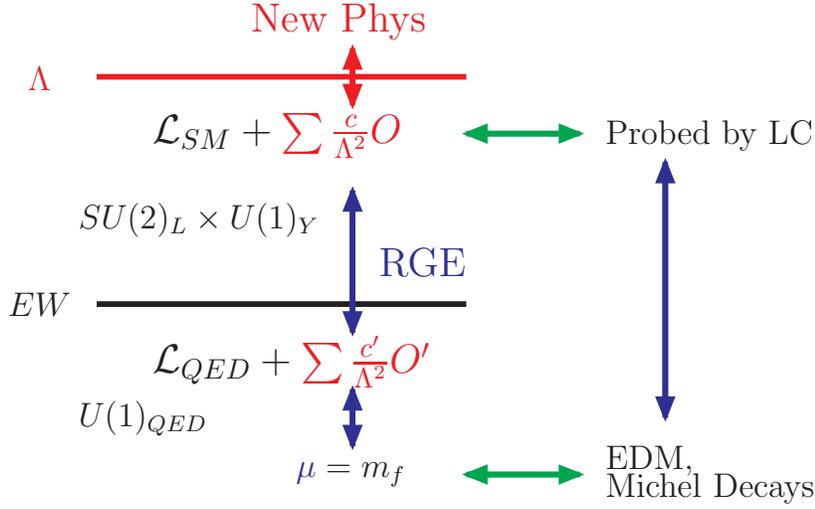, width=0.7\textwidth}
    \end{center}
  \caption{The connection between high and low energy interactions in effective
  field theory. The symmetry at each mass scale is given. All notations are given
  in the text. }
\label{scheme}
\end{figure}

A schematic representation of our approach is given in Fig.(\ref{scheme}).
Since the list of operators
are known for some time; thus in that sense our approach is not new.
However, a detail analysis the CPV effects
incorporating RG effects have not been given before.
\section {Dimension six operators}
Without further ado we give  the most general $SU(2)\times U(1)$ invariant
Lagrangian with dimension six
4-lepton operators. We assume that there are no sterile neutrinos below $\Lambda$.
In the weak basis they are not flavor diagonal and are given below:
\beqa
 {\cal L}^{eff}&=&{\cal L}_{SM}+{\cal L}^6\,, \nonr \\
{\cal L}^6&=& \sum \frac{c_S^{ij,kl}}{\Lambda^2} O_S^{ij,kl}+{\cal L}^6_V + h.c.
\label{Leff}
\eeqa
where
\beq
O_S^{ij, kl}= (\bar{L}_{ia} e_j)( \bar{e}_k L_{la}),
\eeq
and \footnote{ Our notation for $SU(2)$ is slightly different from the
more common one involving Pauli matrices.} the vector like dimension-6
lagrangian is
\beqa
{\cal L}^6_V &=& \frac{c_{LL}^{ij,kl}}{\Lambda^2}(\bar{L}_{ia} \gamma^\mu L_{ja})
(\bar{L}_{kb}\gamma_\mu L_{lb})
+ \frac{c_{LR}^{ij,kl}}{\Lambda^2}(\bar{L}_{ia} \gamma^\mu L_{ja})
(\bar{e}_k\gamma_\mu e_l) \nonr\\
&&+ \frac{c_{RR}^{ij,kl}}{\Lambda^2}(\bar{e}_i \gamma^\mu e_j)
(\bar{e}_k\gamma_\mu e_l)+
\frac{d_{LL}^{ij,kl}}{\Lambda^2}\left(\bar{L}_{ia} \gamma^\mu  L_{jb}\right)
\left( \bar{L}_{kb}\gamma_\mu  L_{la}\right) \nonr \\
&&+ h.c.
\label{OV}
\eeqa
The sum in Eq.(\ref{Leff}) is taken over family indices denoted by  $(i,j,k,l)$
and $SU(2)$ indices are $a,b$.
$L_i, e_j$ are $SU(2)$ lepton doublet and singlet respectively.
All repeated indices are summed unless otherwise specified.
Without knowledge of the new physics generating the operators $O_{S,V}$
the Wilson coefficients $c_{S,V},\ d_{LL}$ are unknown complex parameters.
 There are altogether $324$ parameters and a general analysis is not meaningful.
They contain flavor changing neutral currents at the zeroth order.
From the non-observation of rare muon or $\tau$ decays we learn that these terms
are highly suppressed.
This can be implemented by assuming that the Wilson coefficients of a given
chiral structure denoted generically by $c^{ij,kl}$ is given by
$c^{ij,kl}= (constant)\cdot\delta^{ij}\delta^{kl}$.
This greatly simplifies the analysis and also makes the physics more transparent.
A notable exception to this are  the split fermion models in extra dimension
as discussed in\cite{CNLF}.  With this caveat in mind
we can simplify  Eq.(\ref{OV}) to
\beqa
{\cal L}_V^6&=& \frac{c_{LL}}{\Lambda^2}(\bar{L}_{ia} \gamma^\mu L_{ia})
(\bar{L}_{kb}\gamma_\mu L_{kb})
+ \frac{c_{LR}}{\Lambda^2}(\bar{L}_{ia} \gamma^\mu L_{ia})
(\bar{e}_k\gamma_\mu e_k) \nonr\\
&&+ \frac{c_{RR}}{\Lambda^2}(\bar{e}_i \gamma^\mu e_i)
(\bar{e}_k\gamma_\mu e_k)+
\frac{d_{LL}}{\Lambda^2}\left(\bar{L}_{ia} \gamma^\mu  L_{ib}\right)\left(
\bar{L}_{kb}\gamma_\mu  L_{ka}\right) \nonr \\
&&+ h.c.
\label{OVI}
\eeqa

As for the scalar coefficients in general they are not universal.
The best known examples are
the Yukawa couplings of the SM which are hierarchical and family dependent.
We call this  case type I scalar couplings.
At the opposite end we have the  simplifying case in which
all $c_S$ have the same size will be referred to as type II scalar couplings.
We use these two to
illustrate the very different physics that they represent.
\renewcommand{\theenumi}{\alph{enumi}}
\benu
\item {\bf Type I scalar couplings}\\

Now ${\cal L}^6$ takes the simplified form
\beq
\label{OVII}
{\cal L}^6 = \sum \frac{c_S^{ij,kl}}{\Lambda^2} O_S^{ij,kl}
+ \sum \frac{c_V}{\Lambda^2} O_V^{ij,kl}.
\eeq
where $c_V$ is a generic symbol for Wilson coefficients.
Hermiticity demands that the Wilson coefficients associated with $O_V$ are all real.
Now we can perform
biunitary transformations to the mass eigenbasis of the
charged leptons and neutrinos. It can be seen from Eqs.(\ref{OVI},\ref{OVII})
when they are expanded the
terms associated with the $c$'s are neutral current (NC) terms.
Only $d_{LL}$ involves charged current (CC).
 The NC terms are flavor diagonal and their
 couplings are real. On the other hand  the CC  terms will be
multiplied by the lepton mixing matrix, i.e. the PMNS matrix responsible for
neutrino oscillations \cite{PMNS}.
This is completely analogous to the SM where a phase of the weak charged current
arises from fermion basis rotations and not the gauge coupling which is real.
Next we examine the scalar terms.

The same basis transformation will also multiply the $c_S$'s by the
unitary matrices $U_L$ and $U_R$ which  diagonalize $L_i$ and $e_j$.
Let us denote the mass eigenstates by $L_i^{\prime}$ and $e^{\prime}_j$  respectively.
Then the scalar part of Eq.(\ref{OVII}) becomes
\beq
\label{OSI}
\frac{c_S^{ij,kl}}{\Lambda^2}
\left[\overline{L_p^{\prime}}(U_L^{\dagger})_{pi}(U_R)_{jq}e^{\prime}_q\right]
\left[ \bar{e^{\prime}}_r(U_R^{\dagger})_{rk}(U_L)_{ls}L^{\prime}_s\right]
\eeq
where we have omitted the $SU(2)$ indices and sum over repeated indices.
Although the matrices $U_R$ and $U_L$
are not known beyond that they are unitary we can still extract some general
properties from the above. The effective Lagrangian of Eq.(\ref{OSI})
will lead to effective scalar charged currents as well as flavor changing
neutral currents. Since the couplings
remain complex they will induce TVNC in $\mu$ and $\tau$ leptonic decays
as well as generating EDM for leptons.
Of particular interest are the flavor diagonal NC terms.
 Let us study the term with only the electron whose family
index is 1.
In the mass basis, it is
\beqa
\label{eeterm}
&&\frac{c_S^{ij,kl}}{\Lambda^2}
\left[ \overline{e_L}_1(U_L^{\dagger})_{1i}(U_R)_{j1}e_{R1}\right]
\left[ \overline{e_R}_1(U_R^{\dagger})_{1k}(U_L)_{l1}e_{L1}\right] + h.c. \nonr \\
&&= \frac{c^{\prime 11,11}_S}{\Lambda^2}\overline{e_L}_1e_{R1}\overline{e_R}_1e_{1L}
\eeqa
where we have used the same  labels for mass and weak eigenstates in the above equation.
Clearly the above 4-Fermi operator has a real
coefficient. Similarly one can show that  all terms of the form
$(\overline{l_{Li}}l_{Rj})(\overline{l_{Rj}}l_{Li})$ where
$i,j=e, \mu, \tau$  have real coefficients. Only terms of the
form $(\overline{l_{Li}}l_{Ri})(\overline{l_{Rj}}l_{Lj})$ where
$i\neq j$ can have complex couplings since they
are not hermitian. We now arrive at a general conclusion that besides
the phase in the PMNS matrix low energy
CP violation must involve scalar currents. In  the case of NC effects more than one
family is necessary. This generalizes the expectation from the SM.

To summarize we have shown that in
going to the mass eigenbasis there is no loss of generality.
The effective scalar Lagrangian for the terms
which contain CP phases can be written as
\beq
\label{Lmass}
{\cal {L}}_s =\sum \frac{c_S^{ii,jj}}{\Lambda^2}\overline{L_i}e_{Ri}
\overline{e_{Rj}}L_{j} + h.c.
\eeq
and again for notational simplicity the prime on $c_S$ has been dropped.
As for the vector coupling terms it is easy to see that going to the
mass eigenbasis produces
\beqa
{\cal {L}}_V^6 &=& \frac{c_{LL}}{\Lambda^2}
(\overline{\nu_{iL}}\gamma^{\mu}\nu_{iL}+\overline{e_{iL}}\gamma^{\mu}e_{iL})
(\overline{\nu_{jL}}\gamma_{\mu}\nu_{jL}+\overline{e_{jL}}\gamma_{\mu}e_{jL}) \nonr \\
&+& \frac{c_{LR}}{\Lambda^2}
(\overline{\nu_{iL}}\gamma^{\mu}\nu_{iL}+\overline{e_{iL}}\gamma^{\mu}e_{iL})
( \overline{e_{jR}}\gamma_{\mu}e_{jR}) \nonr \\
&+& \frac{c_{RR}}{\Lambda^2}(\bar{e_{iR}}\gamma^{\mu}e_{iR})
(\bar{e_{jR}}\gamma_{\mu}e_{jR}) \nonr \\
&+& \frac{d_{LL}}{\Lambda^2}\left[(\overline{\nu_{iL}}\gamma^{\mu}\nu_{iL})
(\overline{\nu_{jL}}\gamma_{\mu}\nu_{jL})+
(\overline{e_{iL}}\gamma^{\mu}e_{iL})(\overline{e_{jL}}\gamma_{\mu}e_{jL})
 +2 V_{ij}V^{\dagger}_{kl}(\overline{\nu_{iL}}\gamma^{\mu}e_{jL})
 (\overline{e_{kL}}\gamma_{\mu}\nu_{lL})\right] \nonr \\
&+& h.c.
\label{OVM}
\eeqa
where we are in the mass eigenbasis and $V$ is the PMNS matrix.
For notational simplicity we retain the label of the weak basis.
From the above equation we can see that the only physical phase is in the
$CC \times CC$ terms. The coefficient is a product of the PMNS matrix element
and a complex number $d_{LL}$. On the other hand
 in the $NC\times NC$ channels there are no phases and they also conserve flavor.
We see from
Eqs.(\ref{Lmass})  and (\ref{OVM}) that the rare decay $\mu\ra e\gamma$
is  induced at 2-loops.
This will be reserved for further studies.  Hence our scenario is a very
modest extension of  SM expectations.

\item {\bf Type II scalar couplings}\\
The parameters $c_S^{ij,kl}$ are now independent of the family indices and
can be taken out of the sum. A rotation
to the mass eigenbasis and hermiticity produce only real coupling for $c_S$.
We can now perform a Fierz transformation
and $O_S$ becomes $(\bar{L_i}\gamma^{\mu}L_{i})(\bar{e_j}\gamma_{\mu}e_j)$ and
with only real couplings. This will modify
the overall strengths of the  right-handed lepton NC couplings but it {\it does not}
give rise to TVNC. The strength
of the charged current will be modified as in the previous case.
We shall see later that lepton EDM will also
not be generated up to two loops. Hence, this scenario of extreme simplification is
phenomenologically not interesting and no detail discussion for it will be given.

\eenu

 We note in passing that a possible tensor term
 $O_T = (\bar{L} \sigma^{\alpha\beta} e)(\bar{e} \sigma_{\alpha\beta}L)$
is identically zero independent of the family indices states.
Such terms can only be generated after electroweak
symmetry breaking and are expected to be highly suppressed.
The same is true when $e$ is replaced by a light
sterile neutrino $\nu_r$. However, there are now more operators
similar to ones we are studying that one can construct.
We list them in \cite{sterile}.
Since there is no compelling evidence for light sterile neutrinos we shall
not pursue them further.

Before we proceed we comment on  the other dimension six operators
involving gauge bosons and Higgs fields (see \cite{oplist}).
An example is the operator $O_h =\, \phi \bar{L}\sigma^{\alpha\beta}eW_{\alpha\beta}$ where
$\phi$ is the Higgs doublet and $W_{\alpha\beta}$ is the field strength tensor of $SU(2)$.
These operators are important in discussing electroweak breaking and a detail
study can be found in \cite{Higgsop}.
In most models such an operator is  generated at 1-loop or  higher and thus
more suppressed than the four fermion operators which can usually
be generated at tree level.
Examples can be found in models of extra dimensions \cite{CNLF} where
 $\Lambda$ is given by the mass of Kaluza-Klein modes and left-right symmetric models.
Alternatively one can regard our approach to be the ansatz of 4-Fermi dominance
 below $\Lambda$.

\section{1-loop renormalization of ${\cal L}^6$}
The operators $O_{S,V}$ which we shall generically denote by $O^6_A$ are bare operators.
 Upon renormalization the operators  will mix via
\beq
O^{6}_A=\sum _{B=1}^{5} Z^{-1}_{AB}(Z_L)^{n/2}{Z_e}^{m/2}\, O^6_{Br}
\eeq
where $Z_L$ and $Z_e$ are the wavefunction renormalization constants for the $L$ and
$e$ fields and $n,m$ are the number of such fields in each of the $O_i$  operators.
Hence $n,m = 0, 2, {\mathrm {or}}\ 4$.  The sum  here runs over the five terms
of $O_S$ and $O_V$ and subscript $r$ denotes renormalized quantity.
These five 4-fermi  operators form a complete set at the 1-loop level
in the limit that we can neglect lepton masses.
 The renormalized operator $O^6_{B,r}$ will depend on the t'Hooft renormalization
 scale $\mu$ whereas the bare operators do not. Correspondingly the $\mu$-dependence
 of the Wilson coefficients will be such to render the renormalized effective
 Lagrangian ${\cal L}^6_r$ independent of $\mu$. This leads to the
renormalization group equation for the coefficients in Eqs.(\ref{Leff},\ref{OV}):
\beq
\mu\frac{d}{d\mu}c_A+\sum_B \gamma_{AB}c_B =0
\eeq
where $\gamma_{AB}$ is the anomalous dimension matrix which is non-diagonal.
The values of $c_A$ at the weak scale $v$ are  obtained by solving the above
equation plus boundary conditions which are  the values given at another scale
say at a few TeV where they will  optimistically be measured in the future.
Presently we only have limits on a few $c_A$ at the weak scale taken from
searches at LEP II. We now return to discuss  $\gamma_{AB}$ calculation.

\begin{figure}[h]
     \begin{center}
    \epsfig{file=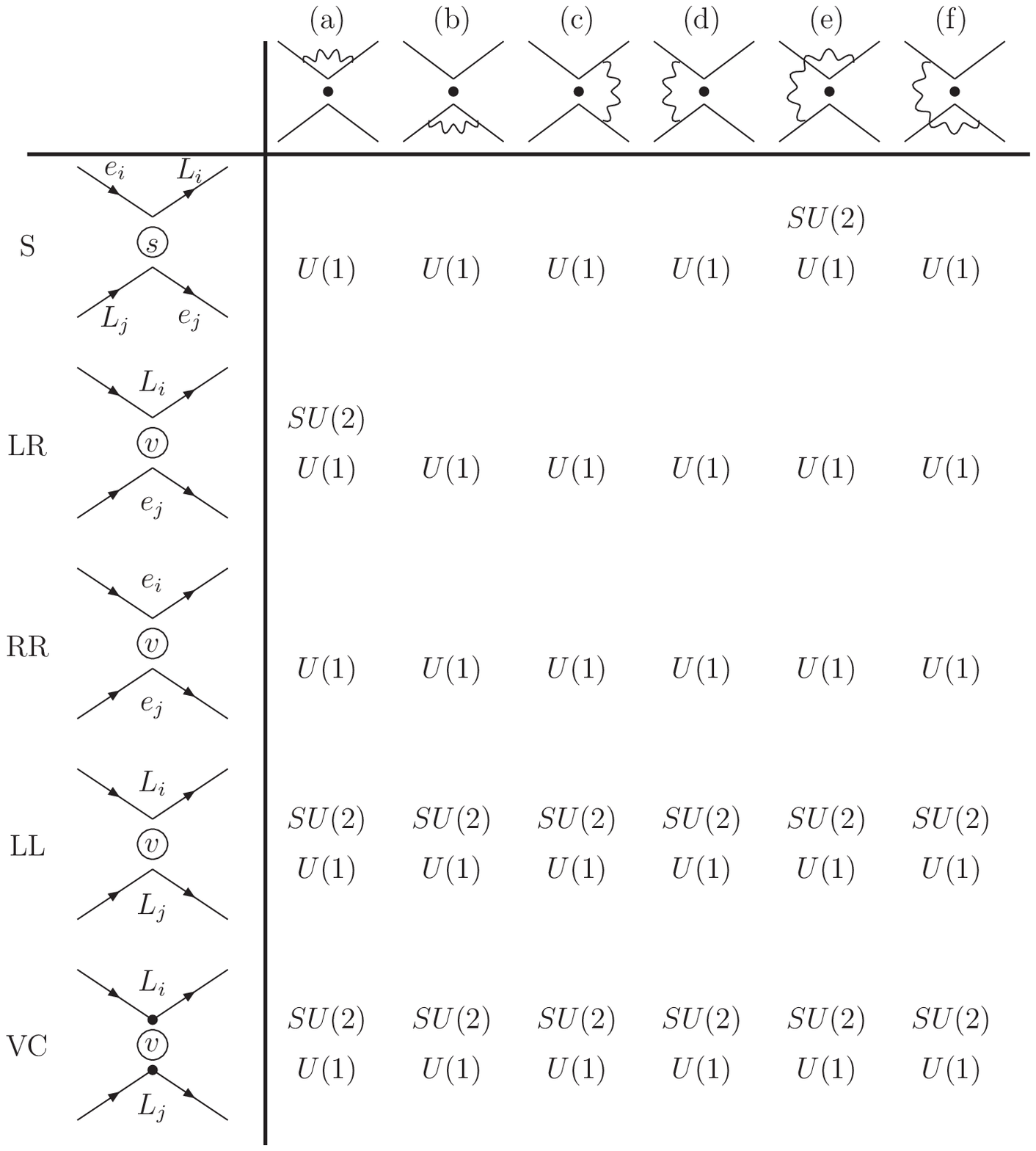, width=0.75\textwidth}
    \end{center}
  \caption{The first column gives the dimension 6 operators. The top
  row depicts their l-loop corrections. The entries in the table are the gauge
   bosons type given by the wavy lines.
   $L$ and $e$ represent the $SU(2)$ doublet and singlet fields respectively. }
\label{fig::lloopmix}
\end{figure}

Between the scales $\Lambda$ and $v$ the 1-loop renormalization of the operators
$O_A$ are induced by the exchange of SM gauge bosons. We can ignore
the Higgs boson contribution  since the Yukawa couplings of the leptons are known to be
small. The task at hand is to calculate the Feynman graphs listed in
 Fig.(\ref{fig::lloopmix}).
Diagrams Fig. \ref{fig::lloopmix}a and \ref{fig::lloopmix}b are cancelled
 by the wave function renormalization graphs (not shown) for the vector coupling case.
 For scalar couplings the
wavefunction renormalizations have to be included. Together with the
remaining four diagram will give the total contribute to the anomalous dimension.

At the 1-loop level operators
of the form $O_h$ are also generated.  Figure (3) depicts two representative diagrams.
\begin{figure}[h]
     \begin{center}
    \epsfig{file=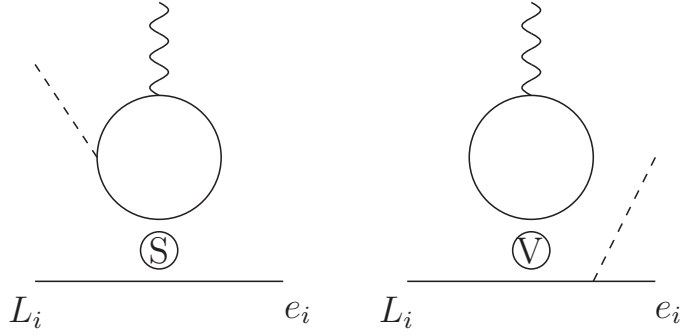, width=0.6\textwidth}
    \end{center}
  \caption{Representative diagrams of dipole operators generated from
  the SM and ${\cal L}^6$. The dash line denotes Higgs fields.
    }
\label{fig::lloopdp}
\end{figure}
Obviously, they are proportional to lepton Yukawa couplings and can be neglected.
In this limit, the four fermion operators, $O_A$,
form a complete set under renormalization to this order.
Thus, the form of Eq.(\ref{Leff}) remains valid at the weak scale.

A standard calculation of the graphs depicted in Fig. 2 gives the following
result for  $\gamma_{AB}$ :
\beq
\gamma_{AB}=\frac{1}{ 4\pi}\left (\begin{array}{ccccc}
-6\alpha_1&0&0&0&0\\
0&-6\alpha_1&0&0&0\\
0&0&12\alpha_1&0&0\\
0&0&0&3(\alpha_1-\alpha_2)&8\alpha_2\\
0&0&0&6\alpha_2&3\alpha_1-7\alpha_2 \end{array} \right )
\eeq
in the basis of operators linked to $(c_S,c_{LR},c_{RR},c_{LL},d_{LL})$
and $\alpha_1$. $\alpha_2$ are the fine structure constants for
the $U(1)$ and $SU(2)$ respectively. This structure holds true for both type I and II
scalar couplings. The difference being that in Type I $c_S$ is a 3$\times$3 matrix
 whereas it is just one parameter for type II.

Below the EW scale the running of the coefficients are governed by QED corrections only.
This is described by the known  $\beta$ functions
for the running of $\alpha_1$ and
$\alpha_2$.

The RGE for $c_S$ and $c_{LR}$ can be solved analytically as they are controlled
by $U(1)$. First we define the quantities
\beq
{\cal G}_{S}(\mu)\equiv { C_S(\mu) \over C_S(M_Z)}\,,\;
{\cal G}_{LR}(\mu)\equiv { C_{LR}(\mu) \over C_{LR}(M_Z)}
\eeq
and obtain the result
\beq
{\cal G}_S(\mu)= {\cal G}_{LR}(\mu) =
 \left( {\alpha_1(M_Z)\over
 \alpha_1(\mu)}\right)^{-\frac{30}{41}}\,.
\eeq
\begin{figure}[h]
    \begin{center}
    \epsfig{file=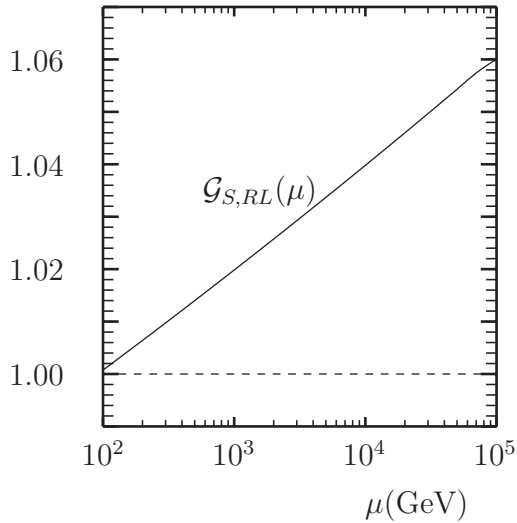, width=0.45\textwidth}
    \end{center}
  \caption{The running of ${\cal G}(\mu)$ defined in the text.}

  \label{fig:RGE_cs}
\end{figure}
As can be seen from  Fig.(\ref{fig:RGE_cs}), numerically the RG effect
is not significant. Even for $\Lambda=10$ TeV it gives
a $\sim 4\%$ correction. The running from weak scale down is also not
 very large and is controlled by the well known
$\beta_{QED}$ \cite{DG}. Including that we find that the Wilson coefficient
for the scalar operator increases by 10\% in going
from the scale of muon mass to 1 TeV.

\section{Lepton EDM from  4-Fermi Interactions}
\begin{figure}[h]
    \begin{center}
    \epsfig{file=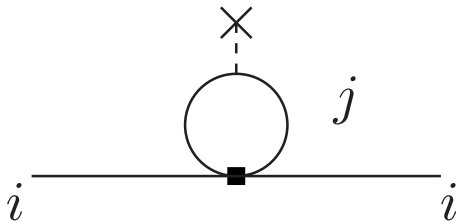, width=0.45\textwidth}
    \end{center}
  \caption{ 1-loop fermion mass from scalar 4-fermion interactions.}
  \label{fig:1lp_mass}
\end{figure}

Next we determine whether ${\cal L}^6$ will generate an EDM at the 1-loop level.
 As a prelude
we  first determine how Eq.(\ref{Lmass}) is related to the lepton masses.
Clearly the 4-fermi
scalar iterations can lead to a mass at 1-loop via
Fig.(\ref{fig:1lp_mass}).
Moreover, in the mass basis there are no off diagonal contributions. Then the
physical mass of a lepton is given by
\beq
\frac{m_i}{v}=y_i +\frac{1}{8\pi^2}\sum_j c_S^{ii,jj}y_j
\label{mdiag}
\eeq
where $y_i$ is the Yukawa coupling of the SM. The first term is the SM contribution.
Ignoring the case of fine tuning between the two terms we expect the real part
of $c_S^{ii,jj}$ to scale like $m_i/m_j$.
In particular for the electron we expect $c_S^{ee,\tau \tau}\sim m_e/m_{\tau}$ and
$c_S^{ee,\mu \mu} \sim m_e/m_{\mu}$. Similarly for the muon.
Notice that for the $\tau$ the main contribution comes from the diagonal coupling
$c_S^{\tau \tau,\tau \tau}$ which can be of order unity.
We note that Eq.(\ref{mdiag}) there are no flavor changing scalar NC in the mass eigenbasis.

\begin{figure}[h]
    \begin{center}
    \epsfig{file=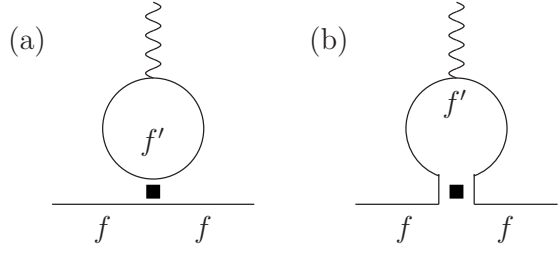, width=0.5\textwidth}
    \end{center}
  \caption{The 2 possible ways for a 4-fermion operator to generate EDM.
  The black box denotes any 4-fermi coupling which is  complex.}
  \label{fig:1lp_EDM}
\end{figure}

Proceeding  to the calculation of EDM,  we observe that at 1-loop there are two
possible ways for a  4-fermion operator to combine with the SM to give rise
 to an EDM, see Fig.\ref{fig:1lp_EDM}.
Now we shall show neither  of them  leads to EDM.
It is more convenient to work in the mass basis.
As seen from Fig.(\ref{fig:1lp_EDM}a) the internal line $f'$ must be a charged lepton.
This narrows the list of contributing operators to the NC type.
We saw in  Sec.2 that the only complex couplings are the ones
associated with the scalar operators $(\overline{f_{L}}f_{R})(\overline{f'_{R}}f'_{L})$
with $f\neq f'$. An elementary calculation shows that Fig.(\ref{fig:1lp_EDM}a)
gives zero contribution. Now we turn to  Fig.(\ref{fig:1lp_EDM}b)
which  does not involve a trace. This comes from operators of the form
$(\overline{f_{L}}f'_{R})(\overline{f'_{R}}f_{L})$. A Fierz transformation
turns this into one associated with $c_{LR}$.
This coefficient is real coefficient and thus not contribute to EDM.
For the case $f=f'$ this Wilson coefficients are real.
Thus, we conclude that there is no 1-loop EDM with purely leptonic
4-fermi operators. We proceed to consider 2 loops.

\begin{figure}[h]
    \begin{center}
    \epsfig{file=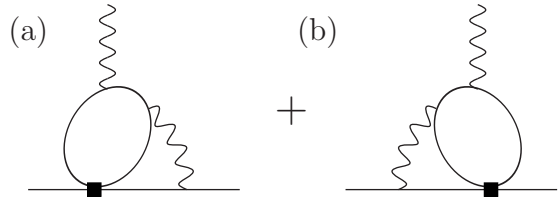, width=0.5\textwidth}
    \end{center}
  \caption{EDM generated by 4-fermion operator at the 2-loop level.}
  \label{fig:2lpEDM}
\end{figure}

The Feynman diagrams we need to consider are  given by
Fig.\ref{fig:2lpEDM}.
Only the coupling $c_S$ represented by the black square can contribute.
The internal wavy line indicates photon exchange and other gauge bosons
exchanges are suppressed. The diagrams are logarithmically divergent due to
the point coupling.
One can regulate this by  substituting $1/\Lambda^2 \rightarrow 1/(q^2-\Lambda^2)$
where $q$ is the loop  momentum carried by the internal photon.
This dampens the high frequency modes of the integration and also reproduces
the point coupling at low energies. Alternatively, one can simply cut off the
integral at $\Lambda$.
We have checked that both ways give the same leading result.
The EDM $d_l$ of a charged lepton $l= e, \mu$ or $\tau$ is given by
\beq
d_f = {e \alpha  \over 16 \pi^3} \sum_i {m_i \mathbf{Im}c^{ff,ii}_S \over \Lambda^2}
{\cal F}\left( {m_i^2 \over \Lambda^2}\right)\,,
\label{EDMresult}
\eeq
 and the function
\beqa
{\cal F}(x)&=& \int^1_0 {  d\lambda\lambda(1-\lambda)\over
x-\lambda(1-\lambda)}\ln\frac{x
}{\lambda(1-\lambda)}\nonr\\
&=& \mbox{Re}\left\{
 {1\over\sqrt{1-4x}}\left[\ln x \ln{\sqrt{1-4x}-1
 \over \sqrt{1-4x}+1 } \right.\right.\\
&+& \left.\left.Li_2\left({2\over 1-\sqrt{1-4x} }\right)
-Li_2\left({2\over 1+\sqrt{1-4x} }\right)
\right]\right\}\,,\nonr\\
{\cal F}(x\ll1 )&\sim& -(2+\ln x)\,,
\eeqa
where $i= e,\mu,\tau$. There are two helicity flips in the above result. The first one
involves the scalar 4-fermi interactions $c_s$ which we have argued before the modulus
of which contains a mass factor $m_i$. The second one is the helicity flip of the
fermion in the loop which is explicit in Eq.(\ref{EDMresult}).
As expected, in the limit of massless fermions  there is no EDM.

The muon and electron EDM's  will be dominated by the diagram with the $\tau$ running
in the loop, and  the result  can be expressed as
\beqa
d_{f}   \sim  5.88 \times 10^{-24}\
\mathbf{Im}c^{ff,\tau\tau}_S \left( { 1 \mbox{TeV }\over \Lambda}\right)^2
(\mbox{e-cm})\,.
\eeqa
As expected  $d_l$ vanishes when the new physics scale becomes arbitrarily
high and if the SM symmetry remains good.

\begin{figure}[h]
  \centering
    \includegraphics[width=0.5\textwidth]{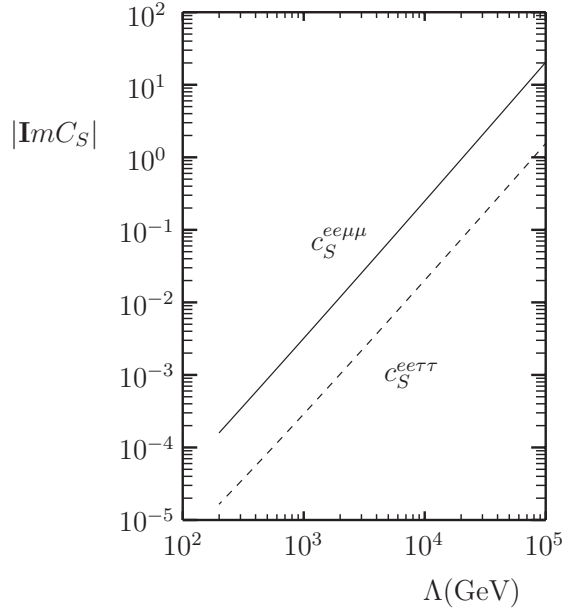}
  \caption{The upper limit of the imaginary part of $c_S$
  derived from the electron EDM. }
  \label{fig:EDM_B}
\end{figure}
For the electron, experimentally we have
$|d_e|< 1.7\times 10^{-27} (\mbox{e-cm})$ \cite{de}
 from which we derive  an upper limit for
 $|\mathbf{Im}c^{ee,\tau\tau}_S|< 3\times 10^{-4}(\Lambda/1
\mbox{TeV})^2$. This in turn can be used to constrain the CP
violating leptonic decays for the $\tau$.
Similarly barring possible cancellations, the electron EDM can also
set an upper bound on the coefficient
$|\mathbf{Im}c^{ee,\mu\mu}_S|< 3.3\times 10^{-3}(\Lambda/1 \mbox{TeV})^2$.
This we shall use  later to limit effects of CP violation in muon decay spectrum.
Figure (\ref{fig:EDM_B}) shows
our result from the electron EDM study.

Similarly, for the $\tau$ EDM, the muon in the loop dominates, and
\beq
d_{\tau} \sim 5.34 \times 10^{-25} \mathbf{Im}c^{\tau\tau,\mu\mu}_S
\left( { 1 \mbox{TeV }\over \Lambda}\right)^2 (\mbox{e-cm})\,.
\eeq

The EDM's can certainly receive a tree level contribution from the dimension six dipole operators :
$\frac{c_{d1}^{ij}}{\Lambda^2} \bar{L}^i\sigma^{\mu \nu} e^j B_{\mu\nu}\phi $ and
$\frac{c_{d2}^{ij}}{\Lambda^2} \bar{L}^i\sigma^{\mu \nu} e^j W_{\mu\nu}\phi $, where $\phi$ is the Higgs
field, and $W_{\mu\nu}$ and $B_{\mu\nu}$ are the $SU(2)$ and $U(1)_Y$  field strength respectively.
After electroweak symmetry breaking, i.e. $\phi\rightarrow v/\sqrt 2\, (0,1)^T$,
this will contribute to EDM and $g-2$ of the lepton at the tree
level. We estimate that the experimental bounds  are
$\left|{\mathbf {Im}}[\cos\theta_W c_{d1}^{ee}+\sin\theta_W c_{d2}^{ee} ]\right| < 2.4\times 10^{-10}$ from $d_e$ and
$\left|{\mathrm {Re}}[\cos\theta_W c_{d1}^{ee}+\sin\theta_W c_{d2}^{ee} ]\right| < 7.1 \times 10^{-6}$ from $a_e$ for $\Lambda = 1$ TeV.
The limit from the muon anomalous moment is about one order of magnitude better.  .
We interpret this to imply that the UV completed theory will either
gives a small phase to this operator and/or it arises from 2 or higher loops with possibly
other dynamical suppression factors.

For completeness we note that the RG running of the dimension-6 electric
dipole operators below the EW scale is not
very significant \cite{DG} and they are not included in this estimate.

\section{More Phenomenological Implications}
\subsection{CP violating Michel parameter in $\mu$  decays}
In the muon decay, $\mu \ra e \nu_\mu \bar{\nu}_e$,
 the electron polarization $\overrightarrow{P_e}$
 is
\beq
\overrightarrow{P_e} =
P_L \hat{z} + P_{T1} {(\hat{z}\times \overrightarrow{P}_\mu)\times \hat{z}
\over |(\hat{z}\times \overrightarrow{P_\mu})\times \hat{z}|}
+P_{T2} {\hat{z}\times \overrightarrow{P}_\mu
\over |\hat{z}\times \overrightarrow{P_\mu}|}
\eeq
where $\overrightarrow{P}_\mu$ is the muon polarization vector
and $\hat{z}$ is the unit vector in the electron momentum
direction.
The $P_{T2}$ term  is a T-violating observable.
To compare with the experimental results
 we follow the convention used by \cite{PDG} to describe
the general decay matrix elements, namely,
\beq
{4G_F \over \sqrt{2}}\sum_{E,M=R,L}^{\gamma=S,V,T}
g^\gamma_{EM}\left\langle \bar{e}_E|\Gamma^\gamma | \nu_e \right\rangle
\left\langle \bar{\nu_\mu} |\Gamma_\gamma | \mu_M \right\rangle
\eeq
where $E,M$ are the chiralities of electron and muon respectively.

The upper limit of  $P_{T2}$  \cite{Burkard}, can be translated into
bounds on two CP-odd parameters $\alpha'= -0.003(69)$
and $\beta'=0.024(101)$ which are
defined as:
\beqa
\alpha'&=& 8 Im\left\{g^V_{LR}(g^{S*}_{RL}+6 g_{RL}^{T*})
- g^V_{RL}(g^{S*}_{LR}+6 g_{LR}^{T*})  \right\}\,,\nonr\\
\beta'&=& 4 Im\left\{g^V_{RR} g^{S*}_{LL}
- g^V_{LL}g^{S*}_{RR} \right\}\,.
\eeqa
The assumption that the SM gauge symmetry is valid up to $\Lambda$ leads to
$g^V_{RR}=g^V_{LR}=g^V_{RL}=0$ automatically. This is consistent with the
latest precision measurement of muon decays \cite{Twist}.
The only new contribution comes from $g^S_{RR}$.
Since the correction is small we can normalize the rate to  $G_F$;
thus, we set $g^V_{LL}=1$. We  predict $\alpha'=0$ and
\beq
\beta' = 4 Im g^S_{RR} = {\sqrt{2} \over G_F \Lambda^2}\mbox{Im}\, c_S^{\mu\mu
ee}\,.
\eeq
From the electron EDM we already have $|\beta'|< 4.0\times 10^{-4}$
for $\Lambda=1$ TeV.
This is two orders of magnitude below the anticipated sensitivity of
the current PSI experiment.

It is interesting to note that $\mbox{Re}\, c_S^{\mu \mu ee}$ is probed by
measuring two other Michel parameters $\eta$ and $\xi$.
The current limit \cite{PDG} translates into
$ |c_S^{\mu\mu ee}|<2.2$ which is not very stringent as compared from the EDM limit.

\subsection{CP violation in $e^+e^- \ra \tau^+\tau^-$ at the Linear Collider}

With the upper bound obtained from EDM we can now estimate the size of CP
violating signatures in a purely leptonic flavor conservation reaction such
as $e^+(p_{+})e^-(p_{-}) \ra \tau^+(k_{+})\tau^-(k_{-})$
where the 4-momentum of each particle are given in the corresponding bracket.
We look for a signal that will be directly sensitive to $Im C_S$.
The signal will involve measuring the  polarization of a final state
$\tau$ in the triple product such as
$(\hat{p}_{-}+\hat{k}_{-})\cdot (\vec{s}_e\times \vec{s}_{\tau})$ where
$\hat{p}_{-}$ is the
unit 3-vector along  the incoming $e^-$ direction, $\hat{k}_{-}$ is the
unit 3-vector along the outgoing $\tau^-$ direction, $\vec{s}_e$ is the
polarization of incoming electron beam, and
$\vec{s}_{\tau}$ is the polarization 3-vector of the $\tau^-$ all in the center
 of mass system.
Usually $\vec{s}_e$ is taken
to be either longitudinally or transversely polarized.
\begin{figure}[h]
    \begin{center}
    \epsfig{file=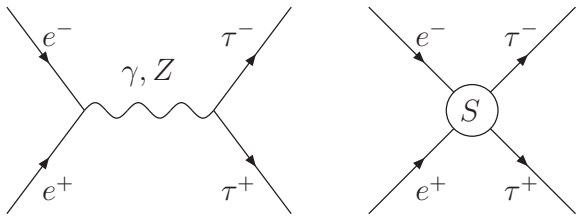, width=0.5\textwidth}
    \end{center}
  \caption{The tree level Feynman diagrams for the CP violating signature in LC. }
  \label{fig:LCFdiag}
\end{figure}
This signature arises from the
interference of an SM amplitude and the 4-fermi term as depicted in Fig.(\ref{fig:LCFdiag}). It is
straightforward to calculate the T-odd (TO) invariant amplitude and we obtain
\beq
|M_{TO}|^2= \frac{s}{24e^2\Lambda^2} Im C_s^{ee,\tau \tau} (\hat{p}_{-}+\hat{k}_{-})\cdot
(\vec{s}_e\times\vec{s}_\tau )
\eeq
where we have scaled by the strength of the QED term and $s$ is the cm energy square.
The linear dependence on $s$ is characteristic of point interactions.
This triple correlation is directly related to EDM measurement.
This can be easily understood since the diagrams of Fig.(\ref{fig:LCFdiag})
are the ones obtained from cutting the 2-loop EDM diagram with amputated photon.
Consequently we obtain an upper bound on the coefficient to be
$10^{-4}(\frac{\sqrt s}{1 \mathrm {TeV}})^2$.
The corresponding quantity using muons will be an order of magnitude higher
due to the less stringent bound on $c_S^{ee,\mu\mu}$ (see Fig. \ref{fig:EDM_B}).

In passing we note that other triple correlations involving two $\tau$ spins
are highly suppressed by $m_e$ and $m_{\tau}$ and will not be a good signature
to probe CPV in contact interactions. This is not the case for $\tau$ weak dipole
moments searches \cite{tauWDM}. The contribution of weak dipole moments are
very small for us. We have also not included the unitarity phase from the SM
which can be calculated precisely including the sign (see \cite{SMuph}).

\section{Conclusions}

Starting from the assumption that the SM gauge symmetry
is valid from the weak scale to some new
physics scale $\Lambda$ and the matter content of the SM we have shown that
the general set of dimension six leptonic 4-fermi operators can be cast into
mass eigenbasis without lost of generality.
This assumption also reduces the number of 4-Fermi operators as compared with
the usual assumption of only $U(1)_{em}$ as the good symmetry.
We then assume that the vector Wilson coefficients of a given
chiral structure are constants. This eliminates zeroth order FCNC.
Then the most interesting non-standard Lorentz structures are the scalar charged and
neutral current interactions.
The overall 4-Fermi coupling will also be renormalized due to
the addition of the $d_{LL}$ term.
There are also modifications of the SM $V,A$ structure to the
neutral currents which can be read from Eq.(\ref{OVM}).
However, due to the high value of  $\Lambda$ these corrections are within
the current bounds.
More interestingly we find that
 the only possible new CP violating terms are the scalar Wilson coefficients
for $(\bar{L_i}e_i)(\bar{e}_j L_j)$ with different family indices $i\ne j$.
We use the renormalization group equations to run the Wilson coefficients at the
scale $\Lambda$ to the weak scale and further down to lower energies of the lepton masses.
We found that there is a 10\% correction to the magnitude of these operators.

The new  phase of the coefficient $c_S^{ii,jj}$ will induce EDM of charged leptons at the
2-loop level. The constraint from the electron EDM is currently the most stringent limit on
$c_S^{ee,jj}/\Lambda^2$ where $j=\mu,\tau$.
This in turn implies that CP violation signature will be too
small to be detected in measurements of the electron spectrum in $\mu$ decays.

At high energies CP violation can occur in $e^+e^- \ra \tau^+ \tau^-$ at the linear
collider by measuring the triple correlation
$(\hat{p}_{-}+\hat{k}_{-})\cdot (\vec{s}_e\times \vec{s}_{\tau})$.
From EDM we estimated the upper limit of this signature to be $\lesssim 10^{-4}$
making this almost impossible to measure.
The case for muons is more promising but still very challenging.
We note that CP violation signatures at colliders are a new way of studying
the phenomenon. However it is an endeavor that requires very high precision at
least in the scenario of 4-fermi operator dominance.

Our analysis is general and independent of the details of the unknown new physics.
We have made very conservative assumptions.
If there are new states found between $\Lambda$ and $v$ our analysis will not apply.
An example will be the existence of sterile neutrinos below $\Lambda$.
As seen from the list given in \cite{sterile} the loop calculation of the
EDM will have to be altered although the RG considerations suffer little change.
On the other hand discovering more Higgs particles will not
alter our analysis below $v$ .
Their effects can be incorporated into the effective scalar couplings.
This study  demonstrates quantitatively the complementarity of high energy and
precision low energy measurements as probes of new physics.
This kind of relations are  quite general and are expected to  exist in many models.
It  can be extended to include 4-Fermi operators with quarks and we will leave this for
a future study.

{\bf Note added}:
After the completion of this work, the paper of Cirigliano {\it el
al} \cite{Cirigliano:2005ck}
was brought to our attention.
The authors used similar effective operator analysis to address the problem of lepton flavor
violation and its connection to neutrino mass generation.
Since they are primarily interested in decays such as $\mu\rightarrow e \gamma$,
the dimension six operators analyzed contain two lepton fields
whereas we analyzed the four lepton terms. The dipole operators play an
important role in their  work in contrast to ours as we are interested in the problem of EDM's.
They also do not consider RG running of the Wilson coefficients.
The importance of a general analysis of dimension six operators respecting gauge symmetry
of the SM in studies of new physics in the lepton is recognized and exploited by both of us in
a different but complementary ways.

\acknowledgments{
This research is supported in part by the Natural Sciences and
Engineering Research Council of Canada.
We are grateful to Prof. Z.C. Ou-Yang and Prof. Y.L. Wu of the
Institute of Theoretical Physics ,
Beijing, for their kind hospitality where part of this work is done.
}


\begin{thebibliography}{99}
\bibitem{sterile}
The list of additional 4-fermi operators with $\nu_R$ are
\beqa
Q_1 &=& \left(\,\overline{e^i_R} \gamma^\mu \nu_R^j\,\right)
\left(\,\overline{\nu_R^k} \gamma_\mu e_R^l\,\right)\,,
\nonr \\
Q_2 &=& \left(\,\overline{L^i} \nu_R^j\,\right)
\left(\,\overline{\nu_R^k} L^l\,\right)\,, \nonr \\
Q_3 &=& \left(\,\overline{L_a^i} e_R^j\,\right)
\left(\,\overline{L^k_b} \nu_R^l\,\right) \epsilon^{ab}\,, \nonr \\
Q_4 &=& \left(\,\overline{L^i_a} \sigma^{\mu\nu} e^j_R\,\right)
\left(\,\overline{L^k_b} \sigma_{\mu\nu} \nu_R^l\,\right)
\epsilon^{ab}\,, \nonr\\
Q_5 &=& \left(\,\overline{\nu_R^i}\gamma^\mu \nu_R^j\,\right)
\left(\,\overline{\nu_R^k}\gamma_\mu \nu_R^l\,\right)\,, \nonr
\eeqa
where $a,b$ are $SU(2)$ indices.
These have to be included when there is evidence for light sterile neutrinos.
Alternatvely one can view the searches for leptonic tensor interactions
at low energies as sterile neutrino searches
since they are necessary for the existence of the above structures.
\end{thebibliography}
\end{document}